\documentclass{article}
\usepackage{ijcai25}
\usepackage[table,xcdraw]{xcolor}  

\usepackage{times}
\usepackage{soul}
\usepackage{url}
\usepackage[hidelinks]{hyperref}
\usepackage[utf8]{inputenc}
\usepackage[small]{caption}
\usepackage{graphicx}
\usepackage{amsmath}
\usepackage{amsthm}
\usepackage{booktabs}
\usepackage{algorithm}
\usepackage{algorithmic}
\usepackage[switch]{lineno}
\usepackage{comment}

\usepackage{booktabs} 

\usepackage{xifthen}

\newcommand{\normal}[3][]{%
    \ifthenelse{\isempty{#1}}
        {\mathcal{N}\left(#2, #3\right)}
        {\mathcal{N}\left(#1|#2, #3\right)}%
}

\usepackage{balance}       
\usepackage[T1]{fontenc}   
\usepackage{txfonts}
\usepackage{mathptmx}
\usepackage{booktabs}
\usepackage{textcomp}
\usepackage{cuted}
\usepackage{multirow}
\usepackage{microtype}        
\usepackage[all]{hypcap}    
\usepackage{ccicons}          
\usepackage[utf8]{inputenc} 
\usepackage{todonotes}
\usepackage{float}        
\usepackage{multirow}     
\usepackage{tabularx}     
\usepackage{booktabs}     
\usepackage{makecell}
\usepackage{changepage}
\usepackage{cuted}
\usepackage{capt-of}
\usepackage{titlesec}
\usepackage[normalem]{ulem}

\title{Human Motion Capture from Loose and Sparse Inertial Sensors with Garment-aware Diffusion Models}

 \author{
 Andela Ilic
 \and
 Jiaxi Jiang\and
 Paul Streli\and
 Xintong Liu\And
 Christian Holz\\
 \affiliations
 Department of Computer Science \\
 ETH Zürich, Switzerland \\[.3em]
{\small\href{https://siplab.org/projects/GarmentInertialPoser}{\color{magenta}{\texttt{https://siplab.org/projects/GarmentInertialPoser}}}}%
\vspace{-10mm}}%

\begin{document}
\maketitle
\vskip -10pt
\begin{abstract}
Motion capture using sparse inertial sensors has shown great promise due to its portability and lack of occlusion issues compared to camera-based tracking. 
Existing approaches typically assume that IMU sensors are tightly attached to the human body. 
However, this assumption often does not hold in real-world scenarios. 
In this paper, we present \emph{Garment Inertial Poser} (GaIP), a method for estimating full-body poses from sparse and loosely attached IMU sensors.
We first simulate IMU recordings using an existing garment-aware human motion dataset.
Our transformer-based diffusion models synthesize loose IMU data and estimate human poses from this challenging loose IMU data.
We also demonstrate that incorporating garment-related parameters during training on loose IMU data effectively maintains expressiveness and enhances the ability to capture variations introduced by looser or tighter garments. 
Our experiments show that our diffusion methods trained on simulated and synthetic data outperform state-of-the-art inertial full-body pose estimators, both quantitatively and qualitatively, opening up a promising direction for future research on motion capture from such realistic sensor placements.

\end{abstract}

\section{Introduction}
\label{sec:intro}

Human pose estimation is essential in healthcare, sports, ergonomics, and AR/VR applications. Traditionally, it has been achieved via images or videos ~\cite{Lan_2023,sosa2023selfsupervised3dhumanpose,pavllo20193dhumanposeestimation,wang2020combining,zhang2022mixste,artacho2020unipose,chen2021anatomy,liu2021graph}, AR/VR devices \cite{zheng2023realisticfullbodytrackingsparse,jiang2022avatarposer,dai2024hmd,millerdurai2024eventego3d} and Inertial Measurement Units (IMUs). In recent years, various models have relied on IMUs as a portable and privacy-friendly solution \cite{yi2024physical,de2024combining}. Some notable examples like PIP \cite{yi2022physicalinertialposerpip}, TIP \cite{Jiang_2022}, TransPose \cite{yi2021transposerealtime3dhuman}, and DynaIP \cite{zhang2024dynamicinertialposerdynaip} focus on full-body pose estimation from a minimal number of IMUs. However, they all share a common limitation: the reliance on IMUs that are tightly strapped to the body, thus sacrificing comfort. In contrast, IMUs loosely attached to garments allow individuals to perform everyday activities while wearing regular clothing without feeling restricted or encumbered. In this way, the tracking of human motion in diverse environments, beyond controlled settings, could be facilitated.

\begin{figure}[t]
    \centering
    \includegraphics[width=1.0\linewidth]{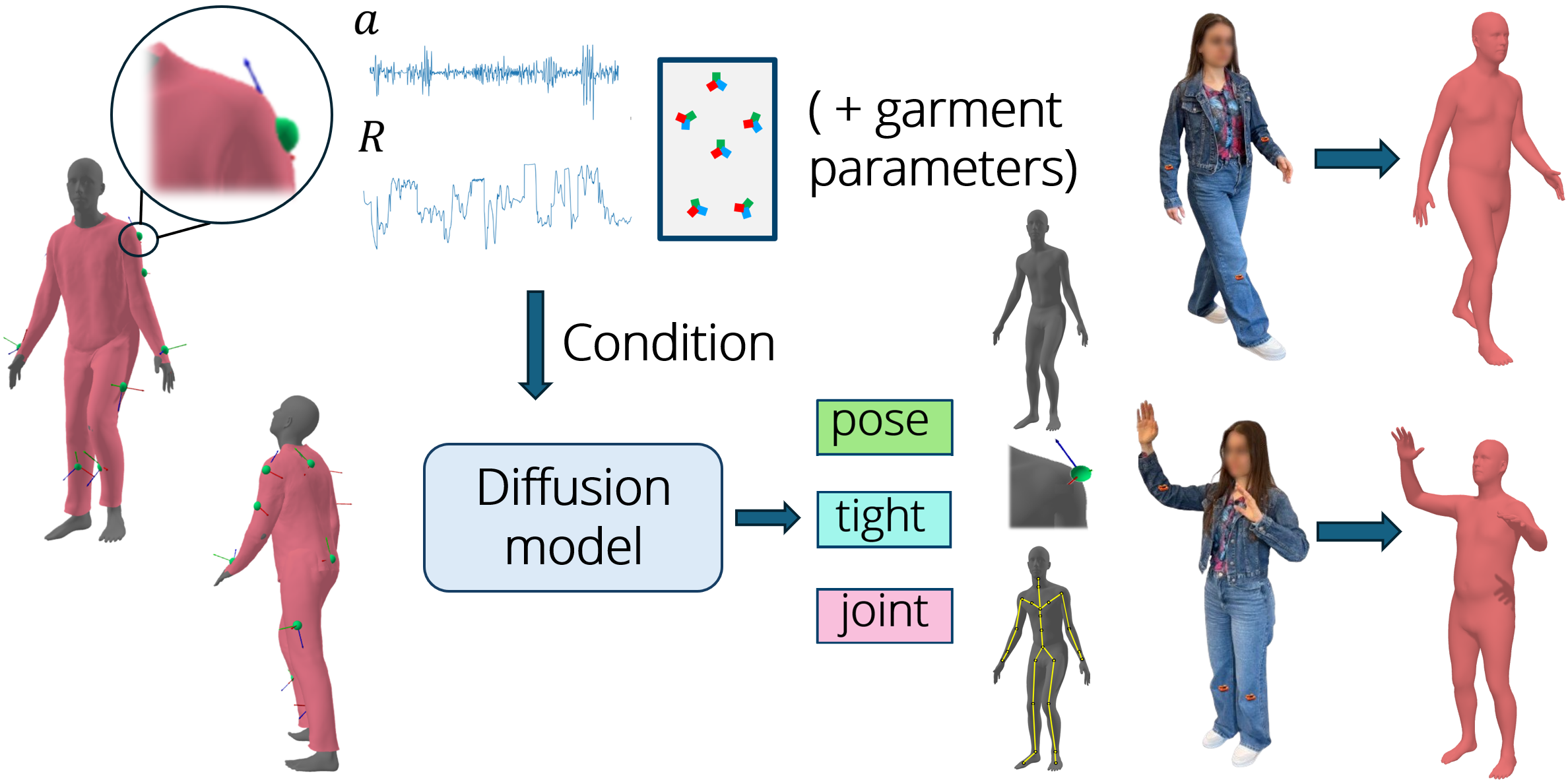}
    \caption{Garment Inertial Poser (GaIP) is a novel model for estimating full-body poses from sparse inertial sensors that are loosely attached to garments. While sensors that are attached to clothing rather than tightly strapped to the body is much more realistic, it is simultaneously more challenging.
    We first introduce diffusion models for simulating and synthesizing loose IMU signals. Leveraging this data, we propose diffusion models conditioned on IMU data alone or IMU data and clothing model parameters, enabling robust human motion estimation.
We evaluate the effectiveness of our approach across three datasets: a simulated dataset, a real-world upper-body dataset, and a real-world full-body dataset.}
    \label{fig:teaser2}
\end{figure}

Loose Inertial Poser (LIP) ~\cite{Zuo_2024_CVPR} explores this setup, focusing on upper-body pose using four IMUs. The Secondary Motion Autoencoder (Semo-AE) they proposed tackles the lack of sufficient real-world training data, as simulated data often fails to capture the full variability seen in actual human movements. However, this model was trained exclusively on recordings from tightly and loosely attached IMUs, disregarding other factors that influence secondary motion, such as the individual’s pose or garment-related characteristics. In addition, it does not address full-body tracking.

To overcome these limitations, we propose a real-time method using conditional diffusion models for full- and upper-body pose estimation from loosely worn IMUs. We train our models on simulated, synthetic, and real datasets, using a secondary diffusion model to generate realistic loose-wear data. While moderately loose conditions represented by simulated and synthetic data provide satisfactory performance, training the model on real-world data reveals a problem with limited motion dynamics and convergence of the model towards the mean prediction. This motivated us to shift from the goal of generating realistic-looking synthetic data to incorporating garment characteristics into the model training. We evaluate our models on simulated~\cite{mahmood2019amassarchivemotioncapture}, real-world upper-body~\cite{Zuo_2024_CVPR}, and real-world whole-body~\cite{ourdataset} datasets, demonstrating improved generalization and robustness to varying garment looseness.

To summarize, our contributions are as follows: 
\begin{enumerate}
    \item the new task of full-body pose estimation using sparse IMU sensors that are loosely attached.
    Previous methods assumed tightly attached IMUs or only attempted partial reconstruction, whereas our work tackles a more challenging and realistic scenario.

    \item 
    a training strategy that applies three distinct types of loosely attached IMU data: simulated data derived from existing garment-aware human motion datasets, synthetic data generated by our secondary diffusion model, and real-world recordings.

    \item 
    a transformer-based diffusion model Garment Inertial Poser (GaIP) that estimates full-body  poses from loosely attached IMUs.
    We show that incorporating garment-related parameters significantly enhances pose accuracy, even amid variations in signal noise and differences in garment fit across individuals.

    \item
    several experimental evaluations that show that our diffusion models outperform the state-of-the-art (SOTA) methods. We also provide comprehensive ablation studies to thoroughly analyze the effectiveness of our methods.

\end{enumerate}

\section{Related Work}

\paragraph{Human pose estimation from wearable sensors.}
Models like DIP, TransPose, TIP, PIP, and DynaIP focus on real-time full-body pose reconstruction using six IMUs. Researchers have noted that this task is under-constrained~\cite{huang2018deepinertialposerlearning,Mollyn_2023}, as many poses can match the same IMU readings. The scarcity of ground-truth data is often addressed by leveraging simulated IMU data derived from mocap datasets. Additionally, modeling temporal dependencies can sometimes impede real-time predictions.
TransPose~\cite{yi2021transposerealtime3dhuman} highlights signal sparsity and noise, while TIP~\cite{Jiang_2022} introduces Transformers to improve temporal consistency and reduce drift. PIP~\cite{yi2022physicalinertialposerpip} is the first real-time, physics-aware model estimating both pose and forces. DynaIP~\cite{zhang2024dynamicinertialposerdynaip} shifts from synthetic to real IMU data and minimizes the influence of less relevant joints to reduce ambiguity.
Beyond standalone IMUs, recent efforts include VR wearables~\cite{jiang2022avatarposer,jiang2024egoposer,jiang2024manikin}, camera-based systems~\cite{hollidt2024egosim}, and hybrid setups combining IMUs with UWB sensors~\cite{armani2024ultra,armani2024icra}.

\paragraph{Pose estimation from loosely attached IMUs.}
To the best of our knowledge, the only method tailored for loosely worn IMUs is Loose Inertial Poser (LIP)~\cite{Zuo_2024_CVPR}. It introduces a Secondary Motion Autoencoder that treats secondary motion as Gaussian noise, using latent space learning to model variation across body shapes and garment types. A temporal coherence mechanism ensures smooth, realistic motion.
Trained on AMASS and evaluated on real upper-body data, LIP achieved $<20$° joint rotation error and 10.6~cm position error and outperformed several SOTA models (PIP, TIP, DIP, TransPose) trained on simulated loose IMU data. LIP uses LSTMs, which suit sequential IMU inputs. In contrast, our pipeline leverages modern generative models to better capture uncertainty and complex motion patterns.

\paragraph{Diffusion models for human motions.}
Diffusion models have recently shown promise for motion generation tasks~\cite{zhang2022motiondiffuse,tevet2022humanmotiondiffusionmodel,tseng2022edgeeditabledancegeneration,yuan2023physdiff,zhou2025emdm,kim2023flame}. In pose estimation, they have been applied to HMD-based setups~\cite{feng2024stratifiedavatargenerationsparse,du2023avatarsgrowlegsgenerating}, where sparse upper-body tracking must be used to infer full-body motion. These models perform well under such under-constrained conditions.
DiffusionPoser~\cite{vanwouwe2024diffusionposerrealtimehumanmotion} is the most relevant work here—it uses IMUs and pressure insoles with an unconditional diffusion model that supports flexible sensor configurations and is robust to signal degradation. Unlike DiffusionPoser, which uses tightly worn sensors, our method employs conditional diffusion models trained on both generated and real-world data with loosely attached IMUs.

\section{Garment Inertial Poser (GaIP) Method}
\label{sec:implementation}
\subsection{Method Overview}

We estimate full-body human pose from loosely attached garment-mounted IMUs using acceleration and orientation inputs to predict SMPL joint rotations.

We generate training data by simulating IMU signals from a garment-aware motion capture dataset. Our approach uses two conditional diffusion models: a primary model for pose estimation and a secondary model for generating realistic loose-wear IMU signals. The secondary model, conditioned on tight-wear IMU data and pose, synthesizes loose-wear signals to train our primary GaIP model. It predicts pose, tight-wear IMU signals, and joint positions. A garment-aware variant further conditions on clothing parameters to enhance realism.

\subsection{Diffusion Framework}

We adopt a Denoising Diffusion Probabilistic Model (DDPM) \cite{ho2020denoisingdiffusionprobabilisticmodels} as the generative backbone. This model operates by progressively corrupting data through a forward process, where Gaussian noise is incrementally added to the data over successive timesteps, forming a series of latent variables \( z_t \). Starting from a clean data sample \( x_0 \), the forward diffusion process follows a Markov chain that applies Gaussian noise at each timestep \( t \), defined as:

\[
q(z_t | x_0) = \mathcal{N}\left(z_t; \sqrt{\alpha_t} x_0, (1 - \alpha_t) \mathbf{I} \right),
\]

where \( \alpha_t \) is a noise scheduling coefficient that controls the amount of noise added at each step, and \( t \in [0, T] \) denotes the progression of time. As \( t \) approaches the maximum timestep \( T \), the sample \( z_T \) approximates a Gaussian distribution \( \mathcal{N}(0, \mathbf{I}) \), effectively turning the original data into pure noise.

To recover the original data, a reverse process is learned, which denoises the latents step by step. The reverse diffusion process is modeled as a sequence of conditional probabilities \( p_\theta(z_{t-1} | z_t) \), where \( \theta \) represents the parameters learned by the model. This process is parameterized by a neural network, which predicts either the noise-free data \( x_0 \) or the noise component \( \epsilon_t \) added during the forward process. For the pose estimation task, we opted to predict the signal itself \( x_0 \), as this facilitates a more straightforward application of the objective function. 

\begin{figure}[b]
    \centering
    \includegraphics[width=\linewidth]{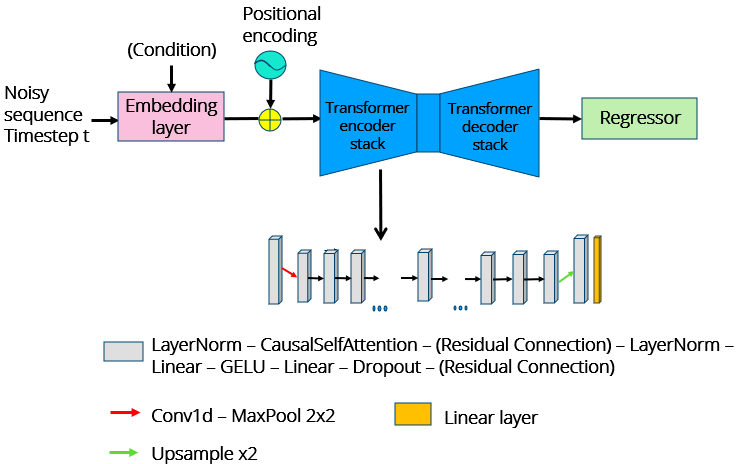}
    \caption{Denoiser architecture used in all diffusion models.}
    \label{fig:denoiser_model}
\end{figure}

Figure \ref{fig:denoiser_model} shows the architecture of the neural network used in all the proposed diffusion models. This is a transformer-based autoencoder with an arbitrary number of transformer blocks in the encoder and decoder networks. 
 
The embedding layer consists of four convolutional layers applied to different parts of the input sequence and the condition, for efficient feature extraction. These parts refer to distinct signals concatenated before being fed into the diffusion model. The use of causal self-attention and residual connections enables effective sequence modeling. To preserve the sequence information, positional encoding is consistently applied before passing the final vector to the first transformer block of the encoder.

\subsection{Data Simulation and Generation}

We simulate both tightly and loosely attached IMU sensor data for poses from the AMASS dataset. The IMU observations include the acceleration and rotation information from 6~sensors:

\[
\text{IMU}(t) = \left( \{a_1(t), a_2(t), \ldots, a_6(t)\}, \{R_1(t), R_2(t), \ldots, R_6(t)\} \right).
\]
The sensors are positioned on the left and right forearm, sternum, pelvis, and left and right lower leg.
For the simulation of clothing effects, we employ TailorNet’s \cite{patel2020tailornetpredictingclothing3d} \textit{Male Shirt} and \textit{Male Pants} models to maintain consistency with the LIP paper. Specifically, we select the SMPL body model \cite{SMPL:2015} and utilize the \textit{TallThin} physique and garment style with zero $\gamma$ parameter.
Following LIP's loose data simulation, we select four neighboring vertices from the clothing model to define the IMU orientation through cross-products, while the geometric center of these vertices serves as its position. Similarly, we obtain tight IMU data directly specifying the corresponding vertex on the AMASS body mesh.

\begin{table}[t]
    \centering
    \resizebox{\columnwidth}{!}{%
    \begin{tabular}{|l|c|c|}
        \hline
        \rowcolor{red!10}
        \textbf{Body Part} & \textbf{Simulated Data [$^\circ$]} & \textbf{Real-World Data [$^\circ$]} \\ 
        \hline
        Left Forearm & 21.451 & 40.890 \\ 
        \hline
        Right Forearm & 21.575 & 49.545 \\
        \hline
        Sternum & 21.094 & 27.395 \\ 
        \hline
        Pelvis & 12.195 & 25.224 \\
        \hline
        Left Lower Leg & 15.819 & 35.449 \\
        \hline
        Right Lower Leg & 13.512 & 36.002 \\
        \hline
    \end{tabular}%
    }
    \caption{Orientation difference between tightly and loosely worn IMU sensors' rotations, calculated as the angular difference in degrees. The difference is computed by converting the rotations to rotation matrices, finding the offset using matrix multiplication, and then applying the axis-angle representation to determine the angle.}
    \label{tab:imu_offsets}
\end{table}

\begin{figure}[b]
    \centering
    \includegraphics[width=0.99\linewidth]{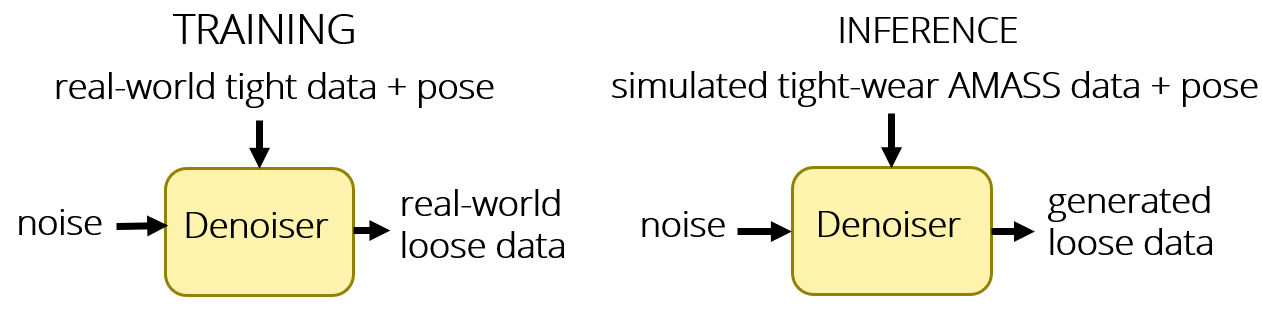}
    \caption{Data generation scheme.}
    \label{fig:data-generation-scheme}
\end{figure}

The orientation difference between the simulated tight and loose data is significantly smaller than that found in real-world data provided by Lorenz et al. \shortcite{ourdataset}, as shown in Table \ref{tab:imu_offsets}. This discrepancy effectively represents tighter-fitting garments, making the simulation less reflective of real-world conditions. To align our dataset better with real-world scenarios, we developed a diffusion model (Figure \ref{fig:data-generation-scheme}) trained on real-world data provided by Lorenz et al. \shortcite{ourdataset}. This model generates loose-wear IMU data \( c_l \) conditioned on tight-wear IMU data \( c_t \) and pose information $\theta$ represented as a quaternion. While the LIP model generates synthetic loose data by introducing noise into the tight data, we believe the pose provides essential secondary motion context. We included pose as the condition, as it directly influences how garments move and interact with the body. The objective function is the simple L1 reconstruction loss:

\begin{equation}
\mathcal{L}_{L1} = \| c_l - \hat{c}_l \|_1 = \sum_{i=1}^{N} | c_l^i - \hat{c}_l^i |.
\end{equation}

Despite the improved offset, we observed that the generated data lacked sufficient diversity, often reflecting scenarios with very wide garment configurations. Therefore, we introduced a parameter $\alpha \sim \text{Uniform}(0, 1)$, which modulates the contribution of the simulated and generated data. By varying 
$\alpha$, we can adjust the tightness or looseness of the clothing representation, effectively blending more or less of the simulated or generated signal. The final training data is obtained using the following formula:

\begin{equation}
c_i = \alpha  c_s + (1 - \alpha)  c_l,
\end{equation}

where \( c_i \) represents the interpolated synthetic loose data, \( c_s \) is the simulated loose data, and \( c_l \) is the loose data generated by our diffusion model.

\subsection{Conditional Diffusion Models for Pose Estimation}

\begin{figure}[htbp]
    \centering
    \includegraphics[width=\linewidth]{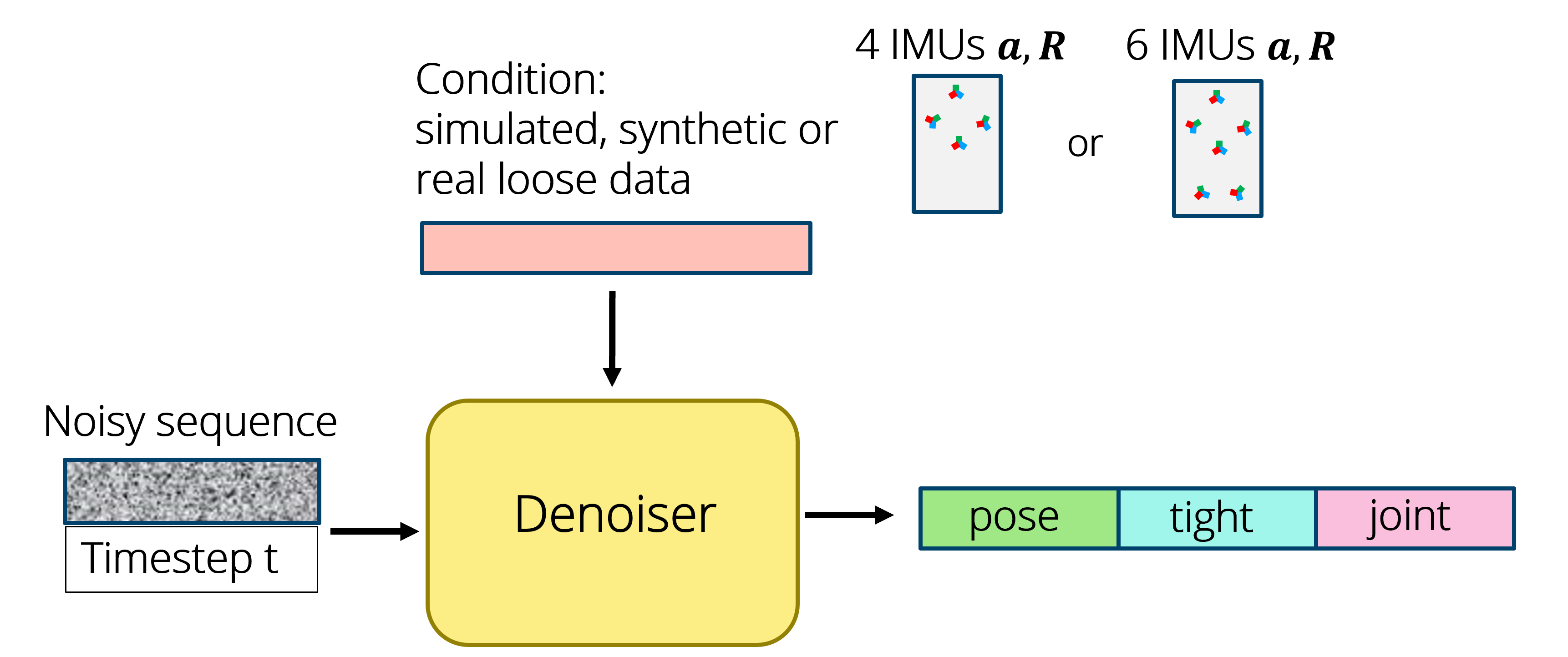}
    \caption{Conditional diffusion model.}
    \label{fig:cond-model-scheme}
\end{figure}

To estimate the pose based on the observations from loosely attached sensors, we developed GaIP, a conditional diffusion model (Figure \ref{fig:cond-model-scheme}). These sensor observations serve as conditions throughout the diffusion process, enabling the model to generate not only pose estimates but also predictions of tight IMU data and joint positions.
Given the conditional nature of this model, where predictions fully rely on the provided input, a flexible number of available sensors cannot be accommodated. Therefore, we propose two distinct models—one for the upper-body pose estimation (GaIP Upper-body) and another for the full-body pose estimation (GaIP Full-body).

GaIP Upper-body predicts the positions and rotations of 14 upper-body joints, including leaf joints, while GaIP Full-body predicts all 24 joints, as defined by the SMPL body model \cite{SMPL:2015}. The loss function for the upper-body model is defined as

\begin{equation}
\begin{split}
\mathcal{L} = \alpha_{r_r} \mathcal{L}_{r_r} + \alpha_{r_j} \mathcal{L}_{r_j} + \alpha_{p_a} \mathcal{L}_{p_a} \\ +  \alpha_{p_j} \mathcal{L}_{p_j} + \alpha_{t} \mathcal{L}_{t} + \alpha_{c} \mathcal{L}_{c}.
\label{eq:loss}
\end{split}
\end{equation}

\vspace{5pt}

Here, \( \mathcal{L}_{r_r} \) and \( \mathcal{L}_{r_j} \) represent the reconstruction errors of the root joint rotation and the remaining joints' rotations, respectively. The terms \( \mathcal{L}_{p_a} \) and \( \mathcal{L}_{p_j} \) capture the position errors for the arms and all other joints, respectively, while \( \mathcal{L}_{t} \) corresponds to the reconstruction error of the tightly attached IMU observations. Finally, \( \mathcal{L}_{c} \) denotes the consistency regularization, calculated as the difference between the model’s predictions when conditioned on the original sensor data and a noisy version of the data, where Gaussian noise scaled by 0.3 is added. The coefficients are set as follows: \( \alpha_{r_r} = 2 \), \( \alpha_{r_j} = 1 \), \( \alpha_{p_a} = 2 \), \( \alpha_{p_j} = 1 \), \( \alpha_{t} = 1 \), and \( \alpha_{c} = 3 \).

The loss function for the whole-body model GaIP Full-body follows the same structure, with \( \mathcal{L}_{p_{al}} \) (position error of arms and legs) replacing \( \mathcal{L}_{p_{a}} \) (position error of arms) to account for the position error of all extremities. Similarly, \( \mathcal{L}_{r_{j}} \) now incorporates the rotation estimation errors for the lower body as well. All loss functions are based on L1 loss. Additionally, we assign higher weights to the root joint’s rotation and the positions of the extremities, as these proved to be particularly challenging for accurate estimation.

\subsection{Garment-Aware Conditional Model}
The conditional model trained on real-world whole-body IMU data revealed issues with very loose garment configurations, particularly with the root joint sensor, which is placed on top of the jacket and over the pants, leading to highly noisy recordings. Similarly, sensors placed on the extremities often provide misleading observations due to garment movement. 
To mitigate this intrinsic uncertainty, we try to enhance GaIP  by integrating garment-related information. To achieve this, we utilize seven simulations generated using TailorNet's \textit{MaleShirt} model. Six of these simulations correspond to a \textit{TallThin} physique, denoted as [180, 22], where 180 represents height in centimeters and 22 refers to the body mass index (BMI). These simulations encompass six garment style variations, with $\gamma$ values of 0, 5, 10, 15, 20, and 24. The final simulation represents a \textit{ShortFat} physique, denoted as [160, 30], with a $\gamma$ value of 0. We focus on four upper-body IMU sensors for these simulations.

The training scheme remains consistent with that shown in Figure \ref{fig:cond-model-scheme}, except for the condition dimension, which now expands to 39 (comprising 36 loose sensor values, $\gamma$, height, and BMI). As a result, the training dataset is expanded by a factor of seven, yielding approximately 47 million samples.

\subsection{Real-Time Inference}

We aim to achieve real-time inference and seamless predictions, which is particularly challenging when working with diffusion models. On one hand, these models require time to denoise the input; on the other hand, they do not keep track of previous estimations. As a result, motion predictions are disconnected, leading to significant jitter.
For all these reasons, we opted for the frame-by-frame prediction strategy. We use a sliding window of length \(N\) (equal to the training sequence length) and shift it by one frame at a time, using the last frame’s prediction as the estimate for the current timestep. To enhance temporal consistency and reduce jitter, we incorporate a masking strategy during inference, as proposed by Van Wouwe et al. \shortcite{vanwouwe2024diffusionposerrealtimehumanmotion}, which suggests fixing past predictions to produce smoother, more coherent movements.
\section{Experiments}
\label{sec:evaluation}
\subsection{Datasets}

\paragraph{AMASS dataset.} AMASS is a large-scale motion capture dataset with more than 40 hours of diverse motions. Specifically, we use the 10 subsets (\textit{BMLrub}, \textit{CMU}, \textit{EyesJapan}, \textit{HumanEva}, \textit{MPIHDM05}, \textit{MPIMoSh}, \textit{SFU}, \textit{SSM}, \textit{Transitions\_mocap}, \textit{BMLmovi}) for training, and the rest 5 subsets \textit{TotalCapture}, \textit{ACCAD}, \textit{TCD\_handMocap}, \textit{BMLhandball}, and \textit{MPI\_Limits}, for the evaluation. These five datasets ensure diversity in motion capture, ranging from everyday activities in \textit{ACCAD} to extreme poses in \textit{MPI\_Limits}, from detailed hand motions in \textit{TotalCapture} and \textit{TCD\_handMocap} to dynamic sports data in \textit{BMLhandball}.

\paragraph{LIP dataset.} The LIP dataset, provided by the authors of the LIP paper \cite{Zuo_2024_CVPR}, contains IMU signals recorded from four sensors placed on the upper body: left forearm, right forearm, back, and waist. The dataset includes two wearing styles: zipped and unzipped, and participants performed walking, running, jumping, boxing and ping-pong, along with five sessions of free-form movement. In total, the dataset comprises 212,496 frames recorded at 30 frames per second.

\paragraph{Whole-body dataset.} Since the LIP dataset is limited to upper-body motions, we utilized a whole-body dataset provided by Lorenz et al. \shortcite{ourdataset}, with 12 IMUs placed on loose-fitting clothing to capture whole-body movements.
Ten participants performed various upper and lower body movements and a 15-minute activity sequence. The pelvis sensor was attached to the jacket, not the trousers, and the work suit sleeves were unbuttoned. The dataset includes 1,137,418 frames recorded at 60 fps.
We utilize only the rotation data from this dataset, since global acceleration measurements are not available.
To obtain the SMPL parameters, we apply inverse kinematics to the global joint orientations provided in the dataset and then map with the corresponding joint names.

\subsection{Evaluation Metrics}
Following the evaluation protocol of LIP~\cite{Zuo_2024_CVPR}, we compute the angular error, which is defined as the Mean Per Joint Rotation Error (MPJRE) in degrees. We also calculate the position error as the Mean Per Joint Position Error (MPJPE) in centimeters. 
To measure the smoothness of the predicted motions, we include two additional metrics: the Mean Per Joint Velocity Error (MPJVE) in $\text{cm/s}$, and Jitter, which quantifies the changes in acceleration over time, expressed in  \(10^2 \, \text{m/s}^3\).
For the upper-body evaluation, we average results across 11 joints, while the whole-body evaluation considers all 24 SMPL model joints.
\subsection{Evaluation Results}
\label{sec:results}
In this section, we compare our GaIP models and the state-of-the-art pose estimation methods using loosely attached IMUs. Note that LIP has previously demonstrated superior performance over methods designed for tightly-attached IMUs \cite{Mollyn_2023,yi2022physicalinertialposerpip,huang2018deepinertialposerlearning,yi2021transposerealtime3dhuman,Jiang_2022}.

Table \ref{tab:performance_metrics} presents the evaluation metrics for the best-performing models across all three datasets. The terms \textit{Sim}, \textit{Syn}, and \textit{Real} in our model names indicate whether the model was trained on simulated, synthetic, or real loose data. 

Table~\ref{tab:comparison_with_tight_models} compares models trained and tested on simulated AMASS data. We evaluate methods designed for loosely attached sensors (Ours, LIP) against those developed for tight-fitting setups (IMUPoser~\cite{Mollyn_2023}, PIP~\cite{yi2022physicalinertialposerpip}) under varying levels of sensor dropout.

\begin{figure}[h]
    \centering
    \includegraphics[width=1.0\linewidth]{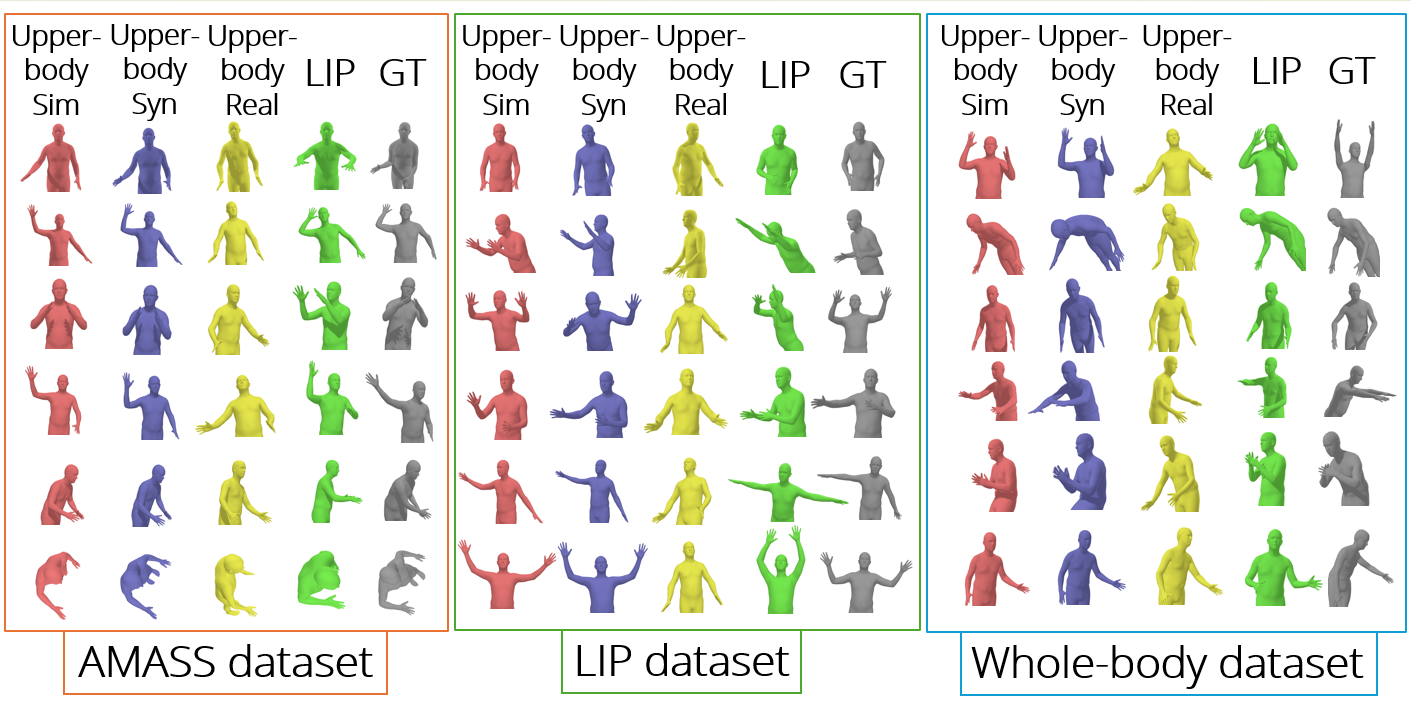}
    \caption{Qualitative performance of the upper-body models on all three evaluation datasets.}
    \label{fig:upper_vis}
\end{figure}

\begin{figure}[htbp]
    \centering
    \includegraphics[width=1.0\linewidth]{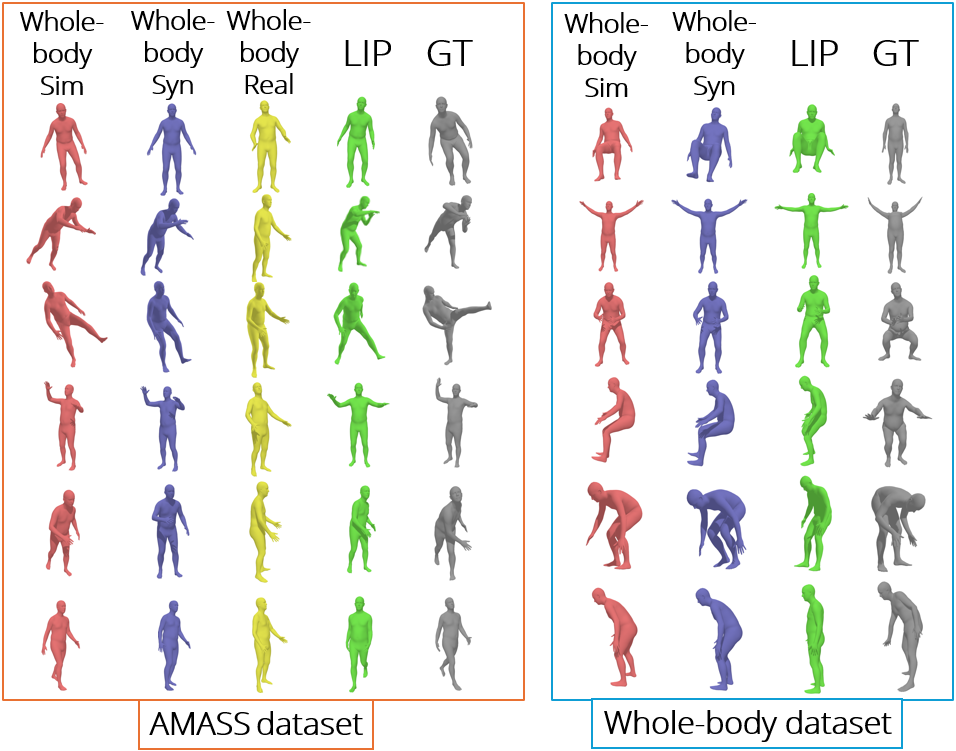}
    \caption{Qualitative performance of the whole-body models on the simulated and real-world datasets.}
    \label{fig:whole_vis}
\end{figure}

\begin{figure}[h]
    \centering
    \includegraphics[width=1.0\linewidth]{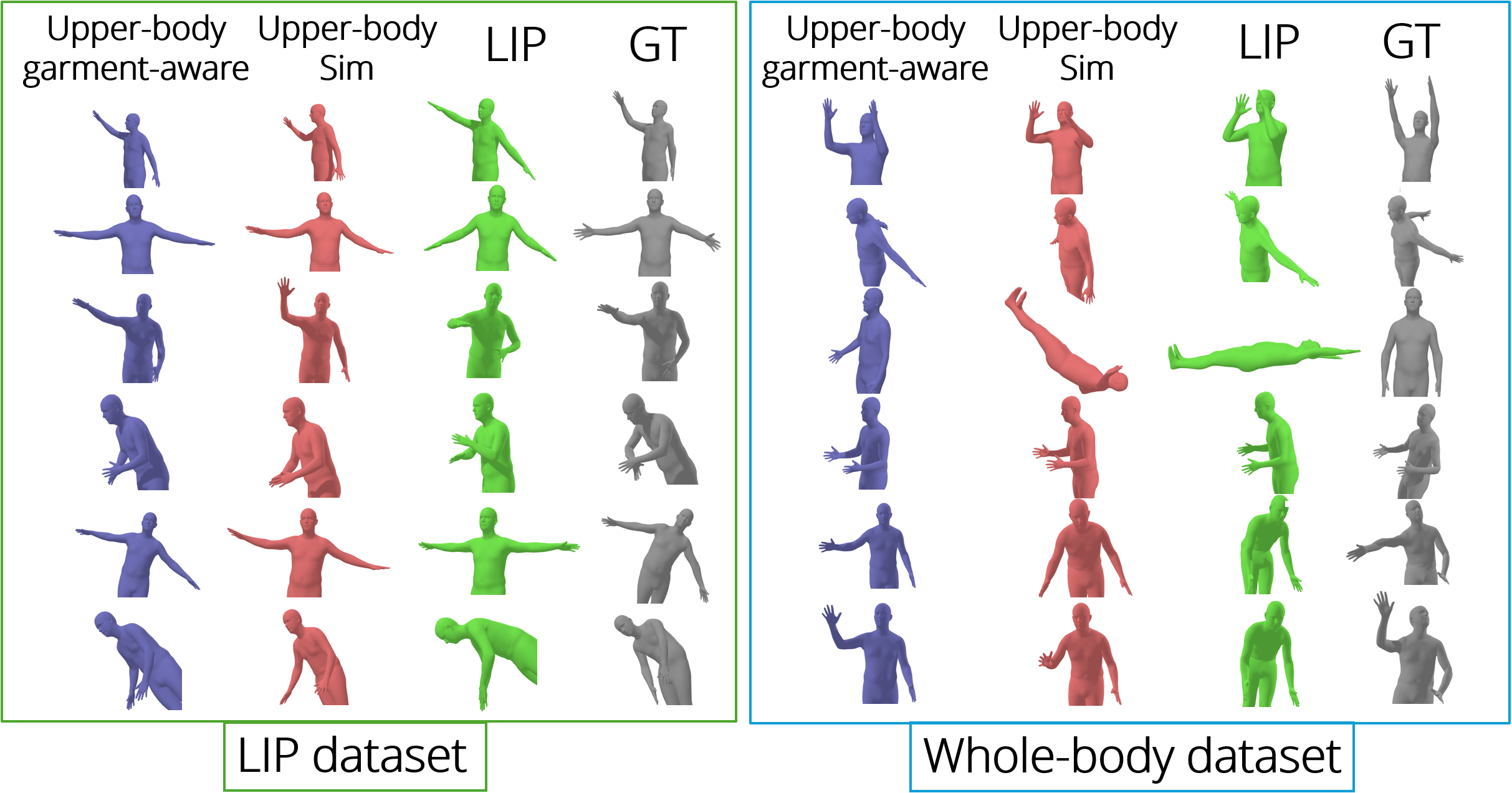}
    \caption{Qualitative performance of the garment-aware conditional upper-body model on the real-world datasets.}
    \label{fig:garment-aware}
\end{figure}

Figures \ref{fig:upper_vis}, \ref{fig:whole_vis} and \ref{fig:garment-aware} illustrate the qualitative performance of our GaIP models, including the upper-body and whole-body garment-unaware models, as well as the upper-body garment-aware model.


\begin{table*}[htbp]
\centering
\normalsize 
\resizebox{\textwidth}{!}{ 
\begin{tabular}{|c|c|c|c|c|c|c|c|c|c|}
\hline
\rowcolor{red!10}
\thead{~~~~ Datasets ~~~~} & \thead{~~~~ Metrics ~~~~} & \thead { ~~~~ Upper-body  ~~~~ \\ Sim (Ours) ~} & \thead{~~~~ Upper-body ~~~~ \\ Syn (Ours) ~} & \thead{~~~~ Upper-body ~~~~ \\ LIP} & \thead{~~~~ Whole-body ~~~~ \\ Sim (Ours) ~} & \thead{~~~~ Whole-body ~~~~ \\ Syn (Ours) ~} & \thead{~~~~ Whole-body ~~~~ \\ LIP ~} & \thead{~~~~ Upper-body ~~~~ \\ Real (Ours) ~} & \thead{~~~~ Whole-body ~~~~ \\ Real (Ours) ~} \\ \hline
\thead{AMASS } & 
\begin{tabular}[c]{@{}c@{}} MPJRE \\ MPJPE \\ MPJVE \\ Jitter \\ GT Jitter \end{tabular} &
\begin{tabular}[c]{@{}c@{}} \textbf{14.46 ± 6.00} \\ \textbf{9.58 ± 6.33} \\ \textbf{26.66} \\ 63.03 \\ 137.07 \end{tabular} &
\begin{tabular}[c]{@{}c@{}} 15.90 ± 6.75 \\ 12.12 ± 9.50 \\ 28.18 \\ 54.65 \\ 137.07 \end{tabular} &
\begin{tabular}[c]{@{}c@{}} 15.05 ± 5.81 \\ 10.58 ± 6.46 \\ 27.65 \\ 95.83 \\ 137.07 \end{tabular} &
\begin{tabular}[c]{@{}c@{}} \textbf{13.65 ± 4.47} \\ \textbf{10.36 ± 5.32} \\ \textbf{27.54} \\ 55.76 \\ 140.76 \end{tabular} &
\begin{tabular}[c]{@{}c@{}} 14.61 ± 4.72 \\ 12.54 ± 6.38 \\ 30.64 \\ 47.00 \\ 140.76 \end{tabular} &
\begin{tabular}[c]{@{}c@{}} 13.67 ± 4.47 \\ 12.78 ± 8.76 \\ 31.40 \\ 93.37 \\ 140.76 \end{tabular} &
\begin{tabular}[c]{@{}c@{}} 27.52 ± 6.89 \\ 15.40 ± 9.57 \\ 31.35 \\ 44.02 \\ 137.07 \end{tabular} &
\begin{tabular}[c]{@{}c@{}} 23.28 ± 4.94 \\ 18.16 ± 9.32 \\ 32.89 \\ 34.34 \\ 140.76 \end{tabular}  \\ \hline
\thead{LIP} & 
\begin{tabular}[c]{@{}c@{}} MPJRE \\ MPJPE \\ MPJVE \\ Jitter \\ GT Jitter \end{tabular} &
\begin{tabular}[c]{@{}c@{}} 20.68 ± 5.66 \\ 12.17 ± 6.01 \\ 32.88 \\ 112.51 \\ 33.41 \end{tabular} &
\begin{tabular}[c]{@{}c@{}} \textbf{19.93 ± 5.68} \\ \textbf{10.65 ± 5.86} \\ \textbf{32.68} \\ 88.40 \\ 33.41 \end{tabular}&
\begin{tabular}[c]{@{}c@{}} 20.85 ± 6.36 \\ 11.49 ± 6.70 \\ 41.99 \\ 233.40 \\ 33.41 \end{tabular} &
\begin{tabular}[c]{@{}c@{}} \_\_ \\ \_\_ \\ \_\_ \\ \_\_ \\ \_\_ \end{tabular} &
\begin{tabular}[c]{@{}c@{}} \_\_ \\ \_\_ \\ \_\_ \\ \_\_ \\ \_\_ \end{tabular} &
\begin{tabular}[c]{@{}c@{}} \_\_ \\ \_\_ \\ \_\_ \\ \_\_ \\ \_\_ 
\end{tabular} &
\begin{tabular}[c]{@{}c@{}} 28.31 ± 5.99 \\ 16.50 ± 7.42 \\ 35.26 \\ 61.89 \\ 33.41 \end{tabular} &
\begin{tabular}[c]{@{}c@{}} \_\_ \\ \_\_ \\ \_\_ \\ \_\_ \\ \_\_ \end{tabular}  \\ \hline
\thead{Whole Body} & 
\begin{tabular}[c]{@{}c@{}} MPJRE \\ MPJPE \\ MPJVE \\ Jitter \\ GT Jitter \end{tabular} &
\begin{tabular}[c]{@{}c@{}} 31.21 ± 7.52 \\ 28.80 ± 16.02 \\ 22.68 \\ 23.61 \\ 43.97 \end{tabular} &
\begin{tabular}[c]{@{}c@{}} \textbf{30.27 ± 8.21} \\ \textbf{25.21 ± 14.10} \\ \textbf{21.12} \\ 15.06 \\ 43.97 \end{tabular} &
\begin{tabular}[c]{@{}c@{}} 33.18 ± 8.84 \\ 26.39 ± 15.59 \\ 23.97 \\ 23.03 \\ 43.97 \end{tabular} &
\begin{tabular}[c]{@{}c@{}} \textbf{24.72 ± 5.70} \\ 28.65 ± 14.47 \\ \textbf{26.97} \\ 34.80 \\ 54.70 \end{tabular} &
\begin{tabular}[c]{@{}c@{}} 25.16 ± 5.95 \\ 28.60 ± 14.37 \\ 28.10 \\ 30.32 \\ 54.70 \end{tabular} &
\begin{tabular}[c]{@{}c@{}} 25.97 ± 7.02 \\ \textbf{26.23 ± 14.5} \\ 30.48 \\ 50.59 \\ 54.70 \end{tabular} &
\begin{tabular}[c]{@{}c@{}} \_\_ \\ \_\_ \\ \_\_ \\ \_\_ \\ \_\_ \end{tabular} &
\begin{tabular}[c]{@{}c@{}} \_\_ \\ \_\_ \\ \_\_ \\ \_\_ \\ \_\_ \end{tabular}  \\ \hline
\end{tabular}
}
\caption{Numerical comparisons between our methods and the state-of-the-art method LIP. We show results for the upper-body as well as the whole-body settings. We also provide the performance of the models trained only on the real-world data to show the benefit of our data simulation and synthesis strategies.}
\label{tab:performance_metrics}
\end{table*}







\begin{table*}[h]
\centering
\scriptsize
\resizebox{\textwidth}{!}{
\begin{tabular}{|c|c|c|c|c|c|c|c|}
\hline
\rowcolor{red!10}
\makecell{Number of \\ missing sensors} & \makecell{~~~~ Metrics ~~~~} & 
\multicolumn{4}{c|}{\cellcolor{red!10} \textbf{AMASS}} & 
\multicolumn{2}{c|}{\cellcolor{red!10} \textbf{Whole-body real-world dataset}} \\ \cline{3-8}
\rowcolor{red!10}
& & \makecell{Whole-body \\ Sim (Ours)} & \makecell{Whole-body \\ LIP} & IMUPoser & PIP & IMUPoser & PIP \\ \hline

\makecell{0} & 
\makecell{MPJRE \\ MPJPE} & 
\makecell{13.65 ± 4.47 \\ 10.36 ± 5.32} & 
\makecell{13.67 ± 4.47 \\ 12.78 ± 8.76} & 
\makecell{12.08 ± 4.77 \\ 7.37 ± 4.43} & 
\makecell{17.38 ± 5.03 \\ 9.62 ± 4.84} & 
\makecell{38.86 ± 5.30 \\ 44.03 ± 8.77} & 
\makecell{43.43 ± 7.27 \\ 44.62 ± 9.53} \\ \hline

\makecell{1} & 
\makecell{MPJRE \\ MPJPE} & 
\makecell{14.02 ± 4.58 \\ 10.71 ± 5.47} & 
\makecell{15.41 ± 4.86 \\ 14.78 ± 8.84} & 
\makecell{12.11 ± 4.86\\ 7.54 ± 4.49} & 
\makecell{20.47 ± 4.35 \\ 23.84 ± 8.51} & 
\makecell{-- \\ --} & 
\makecell{-- \\ --} \\ \hline

\makecell{2} & 
\makecell{MPJRE \\ MPJPE} & 
\makecell{14.56 ± 4.47 \\ 11.76 ± 5.49} & 
\makecell{17.14 ± 4.89 \\ 17.26 ± 9.06} & 
\makecell{12.75 ± 5.04 \\ 8.52 ± 4.76} & 
\makecell{24.84 ± 4.39 \\ 37.82 ± 9.91} & 
\makecell{-- \\ --} & 
\makecell{-- \\ --} \\ \hline

\makecell{3} & 
\makecell{MPJRE \\ MPJPE} & 
\makecell{17.13 ± 4.34 \\ 16.48 ± 6.58} & 
\makecell{19.71 ± 4.62 \\ 21.31 ± 9.19} & 
\makecell{16.05 ± 5.99 \\ 12.26 ± 5.81} & 
\makecell{30.42 ± 4.56 \\ 47.14 ± 9.82} & 
\makecell{-- \\ --} & 
\makecell{-- \\ --} \\ \hline

\end{tabular}
}
\caption{Mean per-joint rotation and position errors on AMASS and a real-world whole-body dataset with varying numbers of missing sensors. Loose-fitting models (Ours, LIP) are compared to tight-fitting ones (IMUPoser, PIP).}
\label{tab:comparison_with_tight_models}
\end{table*}


Based on the results presented in Table \ref{tab:performance_metrics}, we observe that GaIP models trained on simulated and synthetic data perform exceptionally well on the five AMASS datasets. However, their performance gradually declines on the LIP dataset, culminating in significantly reduced accuracy on the whole-body real-world dataset. In contrast, GaIP models trained on real-world data achieve good performance on the whole-body dataset but struggle with the AMASS and LIP datasets, likely due to the tighter fit conditions. This indicates that all GaIP diffusion models and both LIP models perform significantly better when the distribution of test data closely matches that of the training data, revealing a problem with adaptation to different garment conditions or body shapes.

In very challenging scenarios with noisy IMU observations, GaIP models trained on real-world data (Figures \ref{fig:upper_vis} and \ref{fig:whole_vis}) tend to converge towards the mean predictions (neutral pose with legs and arms in relaxed positions). This raises the question of whether the data augmentation approach and emphasis on realistic data are optimal, given that noise and ambiguity in the recorded signals not only mislead the model but also reduce motion dynamics.

We found that incorporating garment-related parameters while training the model on simulated loose data is an effective way to maintain the expressiveness of motion through less noisy observations and still account for potential uncertainties with the addition of these three values. 

While we introduced height and BMI to determine whether a \textit{TallThin} or \textit{ShortFat} physique was used for the clothing simulation, it remains to be seen whether these parameters in real-world scenarios pertain more to garment-related characteristics or the participant's body shape. Since we lacked detailed information about the participants' physiques, we were unable to assess this aspect or evaluate each participant's performance across a long sequence and varying garment styles. However, our experiments on shorter motion sequences, using the optimal selection of the three parameters, demonstrate an improved quantitative performance, as well as a clear qualitative advantage. This is illustrated in Figure \ref{fig:garment-aware}. Both suitable and unsuitable garment-related information can impact pose estimation accuracy, potentially improving or degrading it, with joint angle errors fluctuating by ±5 degrees and joint position errors by up to ±10 cm (compared to the garment-unaware model). Finally, the selection of the three parameters began by choosing values within the training range---height between 160 and 180 cm, BMI between 22 and 30, and the $\gamma$ parameter taking one of six predefined values.
When the parameters were assigned unrealistic real-world height and BMI values, the motion became more restricted and less dynamic, resembling the behavior of the model trained on real-world data with loosely fitting garments (GaIP Upper-body Real).

Experiments with missing sensors (Table~\ref{tab:comparison_with_tight_models}) show that our model handles such conditions better than others. IMUPoser’s performance reflects conservative outputs with limited motion, yielding low error but reduced realism. It also exhibits noticeably more jitter, even when all sensors are available. PIP, on the other hand, performs well only when all sensors are present; its predictions degrade significantly with each additional missing sensor. All models exhibit increased jitter as the number of missing sensors grows\footnote{Visual results are available at \href{https://drive.google.com/file/d/1QHEm2AaGohlv22clJdT9ElAMEb5aAtVc/view?usp=sharing}{https://drive.google.com/ file/d/1QHEm2AaGohlv22clJdT9ElAMEb5aAtVc/view?usp=sharing}.}. The last two columns highlight the superior generalization ability of our models in real-world loose scenarios, where PIP and IMUPoser perform worse on Lorenz et al. \shortcite{ourdataset} dataset.

\subsection{Ablation Study}
\label{sec:ablation}

\paragraph{Consistency regularization loss.}

We attribute GaIP's generalization ability and reduced jitter to the inclusion of this loss function. It compels the model to generate similar outputs for slightly varied and noisier versions of the observations, thereby enhancing its robustness. This is demonstrated by the degraded performance on real-world datasets and the enhanced performance on simulated datasets---which share the same characteristics as the training dataset---when the loss is omitted from the objective function. Table \ref{tab:cr_loss} presents the performance of the upper- and whole-body models trained on simulated data without this loss function.

\begin{table}[t]
\centering
\small
\begin{tabularx}{\columnwidth}{|X|c|c|c|}  
\hline
\rowcolor{red!10}
\thead{Dataset} & \thead{~ Metrics ~} & \thead{Upper-body Sim} & \thead{Whole-body Sim} \\  
\hline

\multirow[t]{5}{=}{\parbox[t]{\linewidth}{\vspace{2mm}\makecell{AMASS \\ dataset}}}  
 & MPJRE & 12.956 ± 5.674 & 12.693 ± 4.573 \\  
 & MPJPE & 7.486 ± 5.255 & 8.130 ± 4.620 \\
 & MPJVE & 24.445 & 25.150 \\
 & Jitter & 130.772 & 106.773 \\
 & GT Jitter & 137.073 & 140.756 \\
\hline

\multirow[t]{5}{=}{\parbox[t]{\linewidth}{\makecell{Real-world \\ upper-body \\ dataset}}}  
 & MPJRE & 21.701 ± 6.390 & \_\_\\  
 & MPJPE & 13.214 ± 7.645 &  \_\_\\
 & MPJVE & 43.738 &  \_\_ \\
 & Jitter & 289.034 &  \_\_\\
 & GT Jitter & 33.412 &  \_\_ \\
\hline

\multirow[t]{5}{=}{\parbox[t]{\linewidth}{\makecell{Real-world \\ whole-body \\ dataset}}}  
 & MPJRE & 32.398 ± 7.362 & 25.758 ± 6.074 \\
 & MPJPE & 27.246 ± 15.289 & 28.647 ± 14.420 \\
 & MPJVE & 23.978 & 29.276 \\
 & Jitter & 50.890 & 72.931 \\
 & GT Jitter & 43.967 & 54.698 \\
\hline

\end{tabularx}
\caption{Performance metrics of GaIP models trained on simulated data without $\mathcal{L}_{\textit{c}}$.}
\label{tab:cr_loss}
\end{table}

\begin{figure}[b]
    \centering
    \includegraphics[width=\linewidth]{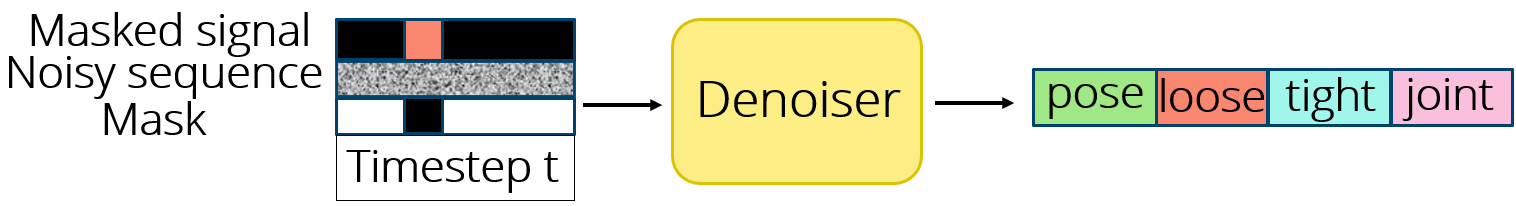}
    \caption{Unconditional inpainting-based diffusion model.}
    \label{fig:uncond-model-scheme}
\end{figure}

\paragraph{Importance of conditioning on all available signals.}
We focused on conditional models that fully rely on provided IMU recordings, a dependency often perceived as their shortcoming. In contrast, an unconditional model can handle missing data, making it particularly useful in scenarios where unreliable signals are received from the root sensor and extremities.  This approach allows for estimating ambiguous signals rather than conditioning predictions on noisy observations. Therefore, we trained an unconditional model to explore this potential improvement. 
During training, we adopted an inpainting method by masking the data from the root sensor, as the root joint estimations of the conditional models exhibited the largest angle errors.
This approach is illustrated in Figure \ref{fig:uncond-model-scheme}.

We generate a mask indicating for each element whether the loose-wear IMU observation is provided (0) or needs to be predicted by the model (1). Following the inpainting approach proposed by Cherel et al. \shortcite{cherel2024diffusionbasedimageinpaintinginternal}, the model input consists of the masked ground truth data \(x_{\text{masked}}\) (Equation \ref{eq:x_masked}), the noisy input, and the generated mask, concatenated along the sequence dimension. Temporal information \( t \) is also included in the input to the denoiser. The model's output is a comprehensive vector that includes pose estimates, loose-wear IMU data, tight-wear IMU data, and joint positions.

\begin{equation}
\mathbf{x}_{\text{masked}} = (1 - \text{mask}) \mathbf{x}_{\text{0}}.
\label{eq:x_masked}
\end{equation}

The objective function mirrors that of GaIP Full-body model, with the addition of the L1 loss term, \( \mathcal{L}_{\text{loose\_recon}} \), to account for errors in the reconstruction of the missing root sensor's data. While including all observations led to similar performance in both the conditional and unconditional models, the results in Table \ref{tab:uncond_results} suggest that completely excluding the root sensor's data from all frames during evaluation is not advisable, as this data is not consistently unreliable.
The presence of reliable observations is likely a contributing factor to the better performance of the model when all data is available.
By omitting certain data entirely, the model is granted the freedom to predict the pose of the affected joint by considering other sensors' data that could be misleading itself (such as data obtained from the extremities), which can lead to inaccurate estimates, particularly pronounced when root rotation is minimal. 

\begin{table}[t]
\centering
\small
\begin{tabular}{|l|c|c|c|c|}
\hline
\rowcolor{red!10}
~
Metrics & Conditional & Unconditional  \\
\hline
MPJRE  &24.722 ± 5.701 & 25.933 ± 5.755 \\
MPJPE  &28.648 ± 14.47& 31.021 ± 11.437 \\
MPJVE  &26.974 & 35.223 \\
Jitter &34.800 &  150.606 \\
GT Jitter &54.698  & 54.698 \\
Root angle error & \textbf{82.123} &  \textbf{103.849} \\
\hline
\end{tabular}
\caption{Comparison between GaIP Full-body model (trained/evaluated on all six sensors) and an inpainting-based unconditional diffusion model (trained/evaluated with missing root sensor).}
\label{tab:uncond_results}
\end{table}

\paragraph{Pose, tightly attached IMU data, and joint positions in output vector.}

We adopted the training scheme presented in Figure \ref{fig:cond-model-scheme}, which involves estimating not only the pose but also data from tightly attached sensors and joint positions. Training the model solely for pose estimation led to abrupt motions. Unlike the loss function in Equation \ref{eq:loss}, which includes six distinct losses, the pose-only model required additional loss components to reduce jitter. These included velocity, acceleration, and jerk losses over one, three, and five frames, as well as velocity losses for joint positions over the same intervals. 
\section{Conclusion}
\label{sec:conclusion}

In this paper, we introduce Garment Inertial Poser---a novel method that estimates full-body pose from sparse and loosely attached IMU sensor data. Our approach builds on a multi-source training strategy that leverages simulated, synthetic, and real-world IMU data. We simulate IMU observations using garment-aware motion datasets and develop a diffusion model to synthesize realistic loose-wear IMU signals. This framework establishes a foundation for a comprehensive comparison of model performance across different data sources, allowing us to systematically evaluate the strengths and limitations of each.
In addition,  we demonstrate that incorporating garment-related parameters during training on simulated loose data enhances model expressiveness and helps address the inherent ambiguities of loosely attached IMU sensors.
Experiments show GaIP outperforms the state-of-the-art methods, providing a new direction for future research. 


\newpage
\bibliographystyle{named}
\bibliography{ijcai25}

\begin{thebibliography}{}

\bibitem[\protect\citeauthoryear{Armani and Holz}{2024}]{armani2024icra}
Rayan Armani and Christian Holz.
\newblock Accurately tracking relative positions of moving trackers based on uwb ranging and inertial sensing without anchors.
\newblock In {\em 2024 IEEE/RSJ International Conference on Intelligent Robots and Systems (IROS)}, pages 12515--12521, 2024.

\bibitem[\protect\citeauthoryear{Armani \bgroup \em et al.\egroup }{2024}]{armani2024ultra}
Rayan Armani, Changlin Qian, Jiaxi Jiang, and Christian Holz.
\newblock Ultra inertial poser: Scalable motion capture and tracking from sparse inertial sensors and ultra-wideband ranging.
\newblock In {\em ACM SIGGRAPH 2024 Conference Papers}, pages 1--11, 2024.

\bibitem[\protect\citeauthoryear{Artacho and Savakis}{2020}]{artacho2020unipose}
Bruno Artacho and Andreas Savakis.
\newblock Unipose: Unified human pose estimation in single images and videos.
\newblock In {\em Proceedings of the IEEE/CVF conference on computer vision and pattern recognition}, pages 7035--7044, 2020.

\bibitem[\protect\citeauthoryear{Chen \bgroup \em et al.\egroup }{2021}]{chen2021anatomy}
Tianlang Chen, Chen Fang, Xiaohui Shen, Yiheng Zhu, Zhili Chen, and Jiebo Luo.
\newblock Anatomy-aware 3d human pose estimation with bone-based pose decomposition.
\newblock {\em IEEE Transactions on Circuits and Systems for Video Technology}, 32(1):198--209, 2021.

\bibitem[\protect\citeauthoryear{Cherel \bgroup \em et al.\egroup }{2024}]{cherel2024diffusionbasedimageinpaintinginternal}
Nicolas Cherel, Andrés Almansa, Yann Gousseau, and Alasdair Newson.
\newblock Diffusion-based image inpainting with internal learning, 2024.

\bibitem[\protect\citeauthoryear{Dai \bgroup \em et al.\egroup }{2024}]{dai2024hmd}
Peng Dai, Yang Zhang, Tao Liu, Zhen Fan, Tianyuan Du, Zhuo Su, Xiaozheng Zheng, and Zeming Li.
\newblock Hmd-poser: On-device real-time human motion tracking from scalable sparse observations.
\newblock In {\em Proceedings of the IEEE/CVF Conference on Computer Vision and Pattern Recognition}, pages 874--884, 2024.

\bibitem[\protect\citeauthoryear{De~Marchi \bgroup \em et al.\egroup }{2024}]{de2024combining}
Mirco De~Marchi, Cristian Turetta, Graziano Pravadelli, and Nicola Bombieri.
\newblock Combining 3d human pose estimation and imu sensors for human identification and tracking in multi-person environments.
\newblock {\em IEEE Sensors Letters}, 2024.

\bibitem[\protect\citeauthoryear{Du \bgroup \em et al.\egroup }{2023}]{du2023avatarsgrowlegsgenerating}
Yuming Du, Robin Kips, Albert Pumarola, Sebastian Starke, Ali Thabet, and Artsiom Sanakoyeu.
\newblock Avatars grow legs: Generating smooth human motion from sparse tracking inputs with diffusion model, 2023.

\bibitem[\protect\citeauthoryear{Feng \bgroup \em et al.\egroup }{2024}]{feng2024stratifiedavatargenerationsparse}
Han Feng, Wenchao Ma, Quankai Gao, Xianwei Zheng, Nan Xue, and Huijuan Xu.
\newblock Stratified avatar generation from sparse observations, 2024.

\bibitem[\protect\citeauthoryear{Ho \bgroup \em et al.\egroup }{2020}]{ho2020denoisingdiffusionprobabilisticmodels}
Jonathan Ho, Ajay Jain, and Pieter Abbeel.
\newblock Denoising diffusion probabilistic models, 2020.

\bibitem[\protect\citeauthoryear{Hollidt \bgroup \em et al.\egroup }{2024}]{hollidt2024egosim}
Dominik Hollidt, Paul Streli, Jiaxi Jiang, Yasaman Haghighi, Changlin Qian, Xintong Liu, and Christian Holz.
\newblock Egosim: An egocentric multi-view simulator and real dataset for body-worn cameras during motion and activity.
\newblock {\em Advances in Neural Information Processing Systems}, 37:106607--106627, 2024.

\bibitem[\protect\citeauthoryear{Huang \bgroup \em et al.\egroup }{2018}]{huang2018deepinertialposerlearning}
Yinghao Huang, Manuel Kaufmann, Emre Aksan, Michael~J. Black, Otmar Hilliges, and Gerard Pons-Moll.
\newblock Deep inertial poser: Learning to reconstruct human pose from sparse inertial measurements in real time, 2018.

\bibitem[\protect\citeauthoryear{Jiang \bgroup \em et al.\egroup }{2022a}]{jiang2022avatarposer}
Jiaxi Jiang, Paul Streli, Huajian Qiu, Andreas Fender, Larissa Laich, Patrick Snape, and Christian Holz.
\newblock Avatarposer: Articulated full-body pose tracking from sparse motion sensing.
\newblock In {\em European conference on computer vision}, pages 443--460. Springer, 2022.

\bibitem[\protect\citeauthoryear{Jiang \bgroup \em et al.\egroup }{2022b}]{Jiang_2022}
Yifeng Jiang, Yuting Ye, Deepak Gopinath, Jungdam Won, Alexander~W. Winkler, and C.~Karen Liu.
\newblock Transformer inertial poser: Real-time human motion reconstruction from sparse imus with simultaneous terrain generation.
\newblock In {\em SIGGRAPH Asia 2022 Conference Papers}, SA ’22. ACM, November 2022.

\bibitem[\protect\citeauthoryear{Jiang \bgroup \em et al.\egroup }{2024a}]{jiang2024manikin}
Jiaxi Jiang, Paul Streli, Xuejing Luo, Christoph Gebhardt, and Christian Holz.
\newblock Manikin: biomechanically accurate neural inverse kinematics for human motion estimation.
\newblock In {\em European Conference on Computer Vision}, pages 128--146. Springer, 2024.

\bibitem[\protect\citeauthoryear{Jiang \bgroup \em et al.\egroup }{2024b}]{jiang2024egoposer}
Jiaxi Jiang, Paul Streli, Manuel Meier, and Christian Holz.
\newblock Egoposer: Robust real-time egocentric pose estimation from sparse and intermittent observations everywhere.
\newblock In {\em European Conference on Computer Vision}, pages 277--294. Springer, 2024.

\bibitem[\protect\citeauthoryear{Kim \bgroup \em et al.\egroup }{2023}]{kim2023flame}
Jihoon Kim, Jiseob Kim, and Sungjoon Choi.
\newblock Flame: Free-form language-based motion synthesis \& editing.
\newblock In {\em Proceedings of the AAAI Conference on Artificial Intelligence}, volume~37, pages 8255--8263, 2023.

\bibitem[\protect\citeauthoryear{Lan \bgroup \em et al.\egroup }{2023}]{Lan_2023}
Gongjin Lan, Yu~Wu, Fei Hu, and Qi~Hao.
\newblock Vision-based human pose estimation via deep learning: A survey.
\newblock {\em IEEE Transactions on Human-Machine Systems}, 53(1):253–268, February 2023.

\bibitem[\protect\citeauthoryear{Liu \bgroup \em et al.\egroup }{2021}]{liu2021graph}
Junfa Liu, Juan Rojas, Yihui Li, Zhijun Liang, Yisheng Guan, Ning Xi, and Haifei Zhu.
\newblock A graph attention spatio-temporal convolutional network for 3d human pose estimation in video.
\newblock In {\em 2021 IEEE international conference on robotics and automation (ICRA)}, pages 3374--3380. IEEE, 2021.

\bibitem[\protect\citeauthoryear{Loper \bgroup \em et al.\egroup }{2015}]{SMPL:2015}
Matthew Loper, Naureen Mahmood, Javier Romero, Gerard Pons-Moll, and Michael~J. Black.
\newblock {SMPL}: A skinned multi-person linear model.
\newblock {\em ACM Trans. Graphics (Proc. SIGGRAPH Asia)}, 34(6):248:1--248:16, October 2015.

\bibitem[\protect\citeauthoryear{Lorenz \bgroup \em et al.\egroup }{2022}]{ourdataset}
Michael Lorenz, Gabriele Bleser-Taetz, Takayuki Akiyama, Takehiro Niikura, Didier Stricker, and Bertram Taetz.
\newblock Towards artefact aware human motion capture using inertial sensors integrated into loose clothing.
\newblock 05 2022.

\bibitem[\protect\citeauthoryear{Mahmood \bgroup \em et al.\egroup }{2019}]{mahmood2019amassarchivemotioncapture}
Naureen Mahmood, Nima Ghorbani, Nikolaus~F. Troje, Gerard Pons-Moll, and Michael~J. Black.
\newblock Amass: Archive of motion capture as surface shapes, 2019.

\bibitem[\protect\citeauthoryear{Millerdurai \bgroup \em et al.\egroup }{2024}]{millerdurai2024eventego3d}
Christen Millerdurai, Hiroyasu Akada, Jian Wang, Diogo Luvizon, Christian Theobalt, and Vladislav Golyanik.
\newblock Eventego3d: 3d human motion capture from egocentric event streams.
\newblock In {\em Proceedings of the IEEE/CVF Conference on Computer Vision and Pattern Recognition}, pages 1186--1195, 2024.

\bibitem[\protect\citeauthoryear{Mollyn \bgroup \em et al.\egroup }{2023}]{Mollyn_2023}
Vimal Mollyn, Riku Arakawa, Mayank Goel, Chris Harrison, and Karan Ahuja.
\newblock Imuposer: Full-body pose estimation using imus in phones, watches, and earbuds.
\newblock In {\em Proceedings of the 2023 CHI Conference on Human Factors in Computing Systems}, volume~38 of {\em CHI ’23}, page 1–12. ACM, April 2023.

\bibitem[\protect\citeauthoryear{Patel \bgroup \em et al.\egroup }{2020}]{patel2020tailornetpredictingclothing3d}
Chaitanya Patel, Zhouyingcheng Liao, and Gerard Pons-Moll.
\newblock Tailornet: Predicting clothing in 3d as a function of human pose, shape and garment style, 2020.

\bibitem[\protect\citeauthoryear{Pavllo \bgroup \em et al.\egroup }{2019}]{pavllo20193dhumanposeestimation}
Dario Pavllo, Christoph Feichtenhofer, David Grangier, and Michael Auli.
\newblock 3d human pose estimation in video with temporal convolutions and semi-supervised training, 2019.

\bibitem[\protect\citeauthoryear{Sosa and Hogg}{2023}]{sosa2023selfsupervised3dhumanpose}
Jose Sosa and David Hogg.
\newblock Self-supervised 3d human pose estimation from a single image, 2023.

\bibitem[\protect\citeauthoryear{Tevet \bgroup \em et al.\egroup }{2022}]{tevet2022humanmotiondiffusionmodel}
Guy Tevet, Sigal Raab, Brian Gordon, Yonatan Shafir, Daniel Cohen-Or, and Amit~H. Bermano.
\newblock Human motion diffusion model, 2022.

\bibitem[\protect\citeauthoryear{Tseng \bgroup \em et al.\egroup }{2022}]{tseng2022edgeeditabledancegeneration}
Jonathan Tseng, Rodrigo Castellon, and C.~Karen Liu.
\newblock Edge: Editable dance generation from music, 2022.

\bibitem[\protect\citeauthoryear{Wang \bgroup \em et al.\egroup }{2020}]{wang2020combining}
Manchen Wang, Joseph Tighe, and Davide Modolo.
\newblock Combining detection and tracking for human pose estimation in videos.
\newblock In {\em Proceedings of the IEEE/CVF Conference on Computer Vision and Pattern Recognition}, pages 11088--11096, 2020.

\bibitem[\protect\citeauthoryear{Wouwe \bgroup \em et al.\egroup }{2024}]{vanwouwe2024diffusionposerrealtimehumanmotion}
Tom~Van Wouwe, Seunghwan Lee, Antoine Falisse, Scott Delp, and C.~Karen Liu.
\newblock Diffusionposer: Real-time human motion reconstruction from arbitrary sparse sensors using autoregressive diffusion, 2024.

\bibitem[\protect\citeauthoryear{Yi \bgroup \em et al.\egroup }{2021}]{yi2021transposerealtime3dhuman}
Xinyu Yi, Yuxiao Zhou, and Feng Xu.
\newblock Transpose: Real-time 3d human translation and pose estimation with six inertial sensors, 2021.

\bibitem[\protect\citeauthoryear{Yi \bgroup \em et al.\egroup }{2022}]{yi2022physicalinertialposerpip}
Xinyu Yi, Yuxiao Zhou, Marc Habermann, Soshi Shimada, Vladislav Golyanik, Christian Theobalt, and Feng Xu.
\newblock Physical inertial poser (pip): Physics-aware real-time human motion tracking from sparse inertial sensors, 2022.

\bibitem[\protect\citeauthoryear{Yi \bgroup \em et al.\egroup }{2024}]{yi2024physical}
Xinyu Yi, Yuxiao Zhou, and Feng Xu.
\newblock Physical non-inertial poser (pnp): Modeling non-inertial effects in sparse-inertial human motion capture.
\newblock In {\em ACM SIGGRAPH 2024 Conference Papers}, pages 1--11, 2024.

\bibitem[\protect\citeauthoryear{Yuan \bgroup \em et al.\egroup }{2023}]{yuan2023physdiff}
Ye~Yuan, Jiaming Song, Umar Iqbal, Arash Vahdat, and Jan Kautz.
\newblock Physdiff: Physics-guided human motion diffusion model.
\newblock In {\em Proceedings of the IEEE/CVF international conference on computer vision}, pages 16010--16021, 2023.

\bibitem[\protect\citeauthoryear{Zhang \bgroup \em et al.\egroup }{2022a}]{zhang2022mixste}
Jinlu Zhang, Zhigang Tu, Jianyu Yang, Yujin Chen, and Junsong Yuan.
\newblock Mixste: Seq2seq mixed spatio-temporal encoder for 3d human pose estimation in video.
\newblock In {\em Proceedings of the IEEE/CVF conference on computer vision and pattern recognition}, pages 13232--13242, 2022.

\bibitem[\protect\citeauthoryear{Zhang \bgroup \em et al.\egroup }{2022b}]{zhang2022motiondiffuse}
Mingyuan Zhang, Zhongang Cai, Liang Pan, Fangzhou Hong, Xinying Guo, Lei Yang, and Ziwei Liu.
\newblock Motiondiffuse: Text-driven human motion generation with diffusion model.
\newblock {\em arXiv preprint arXiv:2208.15001}, 2022.

\bibitem[\protect\citeauthoryear{Zhang \bgroup \em et al.\egroup }{2024}]{zhang2024dynamicinertialposerdynaip}
Yu~Zhang, Songpengcheng Xia, Lei Chu, Jiarui Yang, Qi~Wu, and Ling Pei.
\newblock Dynamic inertial poser (dynaip): Part-based motion dynamics learning for enhanced human pose estimation with sparse inertial sensors, 2024.

\bibitem[\protect\citeauthoryear{Zheng \bgroup \em et al.\egroup }{2023}]{zheng2023realisticfullbodytrackingsparse}
Xiaozheng Zheng, Zhuo Su, Chao Wen, Zhou Xue, and Xiaojie Jin.
\newblock Realistic full-body tracking from sparse observations via joint-level modeling, 2023.

\bibitem[\protect\citeauthoryear{Zhou \bgroup \em et al.\egroup }{2025}]{zhou2025emdm}
Wenyang Zhou, Zhiyang Dou, Zeyu Cao, Zhouyingcheng Liao, Jingbo Wang, Wenjia Wang, Yuan Liu, Taku Komura, Wenping Wang, and Lingjie Liu.
\newblock Emdm: Efficient motion diffusion model for fast and high-quality motion generation.
\newblock In {\em European Conference on Computer Vision}, pages 18--38. Springer, 2025.

\bibitem[\protect\citeauthoryear{Zuo \bgroup \em et al.\egroup }{2024}]{Zuo_2024_CVPR}
Chengxu Zuo, Yiming Wang, Lishuang Zhan, Shihui Guo, Xinyu Yi, Feng Xu, and Yipeng Qin.
\newblock Loose inertial poser: Motion capture with imu-attached loose-wear jacket.
\newblock In {\em Proceedings of the IEEE/CVF Conference on Computer Vision and Pattern Recognition (CVPR)}, pages 2209--2219, June 2024.

\end{thebibliography}

\end{document}